\documentclass[preprintnumbers,color,amsmath,amssymb,floatfix,12pt,prd,superscriptaddress,nofootinbib]{revtex4}
\usepackage{graphicx}
\usepackage{epsfig}       %Revised  on  25 sep  2011%
\usepackage{bm}
\usepackage{amsfonts}
\usepackage{color}
\begin{document}

\title{Holographic dark energy in Brans-Dicke theory with logarithmic correction}

\author{A. Sheykhi}
\email{sheykhi@uk.ac.ir} \affiliation{Department of Physics, Shahid
Bahonar University, P.O. Box 76175, Kerman,
Iran}\affiliation{Research Institute for Astronomy $\&$ Astrophysics
of Maragha (RIAAM), Maragha, Iran}

\author{K. Karami}
\email{KKarami@uok.ac.ir}
 \affiliation{Department of Physics, University
of Kurdistan, Pasdaran St., Sanandaj, Iran } \affiliation{Research
Institute for Astronomy $\&$ Astrophysics of Maragha (RIAAM),
Maragha, Iran}

\author{M. Jamil}
\email{mjamil@camp.nust.edu.pk} \affiliation{Center for Advanced
Mathematics and Physics, National University of Sciences and
Technology, H-12, Islamabad, Pakistan}

\author{E. Kazemi}
\affiliation{Department of Physics, Shahid Bahonar University, P.O.
Box 76175, Kerman, Iran}

\author{M. Haddad}
\affiliation{Department of Physics, Shahid Bahonar University, P.O.
Box 76175, Kerman, Iran}

\begin{abstract}
\vspace*{1.5cm} \centerline{\bf Abstract} \vspace*{1cm} In the
derivation of holographic dark energy density, the area law of the
black hole entropy plays a crucial role. However, the entropy-area
relation can be modified from the inclusion of quantum effects,
motivated from the loop quantum gravity, string theory and black
hole physics. In this paper, we study cosmological implication of
the interacting entropy-corrected holographic dark energy model in
the framework of Brans-Dicke cosmology. We obtain the equation of
state and the deceleration parameters of the entropy-corrected
holographic dark energy in a non-flat Universe. As system's IR
cutoff we choose the radius of the event horizon measured on the
sphere of the horizon, defined as $L=ar(t)$. We find out that when
the entropy-corrected holographic dark energy is combined with the
Brans-Dicke field, the transition from normal state where $w_D >-1 $
to the phantom regime where $w_D <-1 $ for the equation of state of
interacting dark energy can be more easily achieved for than when
resort to the Einstein field equations is
made.\\

%\noindent{\textbf{PACS numbers:}~~~95.36.+x, 04.60.Pp}\\
\end{abstract}

 \maketitle

%%%%%%%%%%%%%%%%%%%%%%%%%%%%%%%%%%%%%%%%%%%%%%%%%%%%%%%%%%%%%%%%%%%%%%%%%%%%%%%%%%%%%%%%%%%%%%%%%%%%%%%%%%%%%%

\newpage
\section{Introduction\label{Int}}
Astrophysical data from type Ia supernovae (SNeIa), cosmic microwave
background radiation (CMB) and large scale structure (LSS) have
provided convincing evidence for the present observable Universe to
be spatially flat and in the phase of accelerated expansion
\cite{perl}. Also most of the portion of cosmic energy density is
contained in the dark sectors i.e. dark energy (DE) and dark matter
(DM) which are 73\% and 23\% respectively while ordinary baryonic
matter (BM) is just 4\%. In the framework of relativistic cosmology,
the cosmic acceleration is described by any perfect fluid whose
pressure $p$ and energy density $\rho$ satisfy $\rho+3p<0$, and such
fluid is termed ``DE'' with negative pressure. In other words, the
equation of state (EoS) parameter $w=p/\rho<-1/3$ theoretically
while observationally it is a daunting task to constrain it. In
theory, there are numerous candidates to explain DE including
cosmological constant, quintessence, phantom energy, K-essence,
quintom, Chaplygin gas, tachyon and modified gravity, to name a few
(see \cite{sami} for comprehensive reviews on DE).

In literature, we have another candidate of DE namely holographic DE
(HDE) which is motivated from the ``holographic principle''
\cite{li,Huang,sus,sus1}. It was shown in \cite{cohen} that in
quantum field theory, the UV cutoff $\Lambda$ should be related to
the IR cutoff $L$ due to limit set by forming a black hole. If
$\rho_D=\Lambda^4$ is the vacuum energy density caused by UV cutoff,
the total energy of size $L$ should not exceed the mass of the
system-size black hole:
\begin{equation}
E_D\leq E_{BH}\rightarrow L^3\rho_D\leq M_p^2L,\nonumber
\end{equation}
where $M_p$ is the reduced Planck mass $M_p^{-2}=8\pi G$. If the
largest cutoff $L$ is taken for saturating this inequality, we get
the energy density of HDE as
\begin{equation}
\rho_D=3c^2M_p^2L^{-2},\nonumber
\end{equation}
where $c^2$ is a dimensionless constant. Following Guberina et al.
\cite{Guberina}, there is an alternative derivation of HDE based on
the entropy bound. In the thermodynamics of the black hole
\cite{Bekenstein}, there is a maximum entropy in a box of size $L$,
namely, the Bekenstein-Hawking entropy bound $S_{BH}\sim M_p^2 L^2$,
which scales as the area of the box $A \sim L^2$, rather than the
volume $V \sim L^3$. Also for a macroscopic system in which
self-gravitation effects can be disregarded, the Bekenstein entropy
bound $S_B$ is given by the product of the energy $E\sim\rho_DL^3$
and the linear size $L$ of the system. Requiring $S_B\leq S_{BH}$,
namely $EL\lesssim M_p^2 L^2$, one has the same result
$\rho_D\lesssim M_p^2L^{-2}$ obtained from energy bound argument.

The HDE is thoroughly investigated in the literature in various
ways. In \cite{inflation}, the HDE is used to drive inflation in the
early Universe. In \cite{jamil}, the EoS of HDE is studied with
varying Newton's gravitational constant and is shown that the EoS
parameter can be modified significantly in the low redshift limit.
In other papers \cite{Karami,Sheykhi}, the HDE is investigated with
different IR cutoffs like the particle horizon, Hubble horizon,
future event horizon and the Granda-Oliveros cutoff. Similarly,
correspondences are established between HDE and other scalar field
models of DE \cite{fehri} while in other studies, HDE is studied in
alternative gravity theories like Braneworld, $f(R)$, scalar-tensor
gravity, Brans-Dicke (BD) and DGP model etc \cite{nozari}. The HDE
also best fits with the observational data of CMB and supernova of
type Ia \cite{observe}.

We emphasize that the black hole entropy $S$ plays a central role
in the derivation of HDE density. Indeed, the definition and
derivation of HDE density depends on the entropy-area relationship
$S\sim
 A \sim L^2$ (or in general $S(A)$) of black holes in Einstein's gravity,
  where $A \sim L^2$ denotes
the area of the black hole horizon. However, this definition can
be modified from the inclusion of quantum effects, motivated from
the loop quantum gravity (LQG). These quantum corrections provided
to the entropy-area relationship leads to the curvature correction
in the Einstein-Hilbert action and vice versa \cite{Zhu}. The
corrected entropy takes the form \cite{modak}
\begin{equation}\label{S}
S=\frac{A}{4G}+\tilde{\alpha} \ln {\frac{A}{4G}}+\tilde{\beta},
\end{equation}
where $\tilde{\alpha}$ and $\tilde{\beta}$ are dimensionless
constants of order unity. The exact values of these constants are
not yet determined and still an open issue in quantum gravity. These
corrections arise in the black hole entropy in Loop Quantum Gravity
(LQG) due to thermal and quantum fluctuations \cite{Rovelli}.
Moreover in Wald's approach to classical gravity and in string
theory, one can find similar corrections to entropy \cite{wald}.
Generally the entropy-area relation can be expanded in an infinite
series expression, but the contribution of these extra terms to the
black hole entropy is negligible due to smallness of $\hbar$
\cite{Zhu}. Hence the leading order term in the expansion is the
logarithmic term to entropy-area relation as considered in Eq.
(\ref{S}). The logarithmic term also appears in a model of entropic
cosmology which unifies the early-time inflation and late-time
cosmic acceleration of the Universe \cite{Liu}. Taking the corrected
entropy-area relation (\ref{S}) into account, and following the
derivation of HDE (especially the one shown in \cite{Guberina}), the
energy density of the HDE will be modified as well. On this basis,
Wei \cite{wei} proposed
 the energy density of the so-called ``entropy-corrected HDE''
 (ECHDE) in the form
\begin{equation}\label{rhoS}
\rho _{D }=3c^2M_{p}^{2}L^{-2}+\alpha L^{-4}\ln
(M_{p}^{2}L^{2})+\beta L^{-4},
\end{equation}
where $\alpha$ and $\beta$ are dimensionless constants of order
unity. In the special case  $\alpha=\beta=0$, the above equation
yields the well-known HDE density. Since the last two terms in Eq.
(\ref{rhoS}) can be comparable to the first term only when $L$ is
very small, the corrections make sense only at the early stage of
the Universe. When the Universe becomes large, ECHDE reduces to the
ordinary HDE.

In particular, the HDE has been widely analyzed in the framework of
BD gravity \cite{kim1,BD,Sheykhi1}. Since the HDE density belongs to
a dynamical cosmological constant, we need a dynamical frame to
accommodate it instead of general relativity. Further, taking $L=
H^{-1}$, it fails to determine the EoS $w_{D}$ in the general
relativity framework. In addition to these, the BD scalar field
speeds up the expansion rate of a dust matter dominated era (reduces
deceleration), while slows down the expansion rate of cosmological
constant era (reduces acceleration). Since our paper deals with the
ECHDE, we generalize the above studies.

In the light of all mentioned above, the investigation on the HDE
models in the framework of BD theory is well motivated. In these
studies \cite{kim1,BD,Sheykhi1}, several dynamical features of HDE
have been explored in the flat/non-flat FRW background e.g. the
phantom crossing ($w=-1$) at the present time; cosmic-coincidence
problem; effective EoS; the deceleration parameter and the quintom
behavior.

This paper is outlined as follows. In section II we study ECHDE in
the framework of BD theory in a non-flat Universe. We also discuss
some of the features of this model including effective EoS,
deceleration parameter and evolution of dimensionless energy density
in the absence of interaction between DE and DM in section II. In
section III, we extend our study to the case where there is an
interaction between ECHDE and DM. The last section is devoted to
conclusions.

\section{ECHDE in BD theory\label{ECHDE}}
The BD action is given by
\begin{equation}
 I=\int{
d^{4}x\sqrt{g}\left(-\varphi {R}+\frac{\omega}{\varphi}g^{\mu
\nu}\partial_{\mu}\varphi \partial_{\nu}\varphi +L_M
\right)}.\label{act0}
\end{equation}
Using the following definition
\begin{equation}
 \varphi=\frac{\phi^2}{8\omega},
\end{equation}
the above action can be rewritten in the canonical form
\cite{Arik}
\begin{equation}
I=\int{ d^{4}x\sqrt{g}\left(-\frac{1}{8\omega}\phi ^2
{R}+\frac{1}{2}g^{\mu \nu}\partial_{\mu}\phi \partial_{\nu}\phi +L_M
\right)},\label{act1}
\end{equation}
where $g$, $\omega$, $\phi$, $R$, and $L_{M}$ are the determinant of
the metric $g^{\mu\nu}$ of spacetime, the BD parameter, the BD
scalar field, the scalar curvature, and the lagrangian of the
matter, respectively. The non-minimal coupling term $\phi^2R$
replaces with the Einstein-Hilbert term $R/G$ in such a way that
$G^{-1}_{\rm eff}= 2\pi\phi^2/\omega$, where $G_{\rm eff}$ is the
effective gravitational constant as long as the dynamical scalar
field $\phi$ varies slowly.

We consider the Friedmann-Robertson-Walker (FRW) metric for the
non-flat Universe as
\begin{eqnarray}
 ds^2=dt^2-a^2(t)\left(\frac{dr^2}{1-kr^2}+r^2d\Omega^2\right),\label{metric}
\end{eqnarray}
where $k=0,1,-1$ represent a flat, closed and open FRW Universe,
respectively. Observational evidences support the existence of a
closed Universe with a small positive curvature ($\Omega_{k}\sim
0.02$) \cite{Bennett}.

Taking the variation of the action (\ref{act1}) with respect to the
metric (\ref{metric}), one can obtain the field equations for the
non-flat Universe containing DE and pressureless dust matter as
\begin{eqnarray}
 &&\frac{3}{4\omega}\phi^2\left(H^2+\frac{k}{a^2}\right)-\frac{1}{2}\dot{\phi} ^2+\frac{3}{2\omega}H
 \dot{\phi}\phi=\rho_D+\rho_M,\label{FE1}\\
 &&\frac{-1}{4\omega}\phi^2\left(2\frac{{\ddot{a}}}{a}+H^2+\frac{k}{a^2}\right)-\frac{1}{\omega}H \dot{\phi}\phi -\frac{1}{2\omega}
 \ddot{\phi}\phi-\frac{1}{2}\left(1+\frac{1}{\omega}\right)\dot{\phi}^2=p_D,\label{FE2}\\
 &&\ddot{\phi}+3H
 \dot{\phi}-\frac{3}{2\omega}\left(\frac{{\ddot{a}}}{a}+H^2+\frac{k}{a^2}\right)\phi=0,
 \label{FE3}
\end{eqnarray}
where $\rho_D$ and $p_D$ are the DE density and pressure,
respectively. Also $\rho_M=\rho_{DM}+\rho_{BM}$ is the total
energy density of pressureless DM and BM. We neglect the
contribution of radiation.

At this point our system of Eqs. (\ref{FE1})-(\ref{FE3}) is not
closed and we still have freedom to choose one. We shall assume that
BD field can be described as a power law of the scale factor,
$\phi\propto a^{n}$. In principle there is no compelling reason for
this choice. However, it has been shown that for small $n$ it leads
to consistent results \cite{Pavon2,Xu}. A case of particular
interest is that when $n$ is small whereas $\omega$ is high so that
the product $n\omega$ results of order unity \cite{Pavon2}. This is
interesting because local astronomical experiments set a very high
lower bound on $\omega$; in particular, the Cassini experiment
implies that $\omega>10^4$ \cite{Bert}. Taking the derivative with
respect to time of relation $\phi\propto a^n$, we get
\begin{eqnarray}\label{dotphi}
&&\dot{\phi}=nH\phi, \\
&&\ddot{\phi}=n^2H^2\phi+n\phi\dot{H}.\label{ddotphi}
\end{eqnarray}

In the framework of BD cosmology, we write down the energy density
of the ECHDE model in the Universe as
\begin{equation}\label{rho1}
\rho_{D}= \frac{3c^2\phi^2 }{4\omega
L^2}+\frac{\alpha}{L^4}\ln\left(\frac{{\phi^2}{L^2}}{4
\omega}\right)+\frac{\beta}{L^4},
\end{equation}
which can be rewritten as
\begin{equation}
\rho_D=\frac{3c^2\phi^2}{4\omega L^2}\gamma_c,\label{ecbd2}
\end{equation}
where
\begin{equation}
\gamma_c=1+\frac{4\omega\alpha}{3c^2\phi^2L^2}\ln
\left(\frac{\phi^2L^2}{4\omega}\right)+\frac{4\omega\beta}{3c^2\phi^2L^2}.\label{gamma}
\end{equation}
For $\alpha=\beta=0$ we have $(\gamma_c=1)$ and thus
\begin{equation}\label{rho1 HDE in B-D}
\rho_{D}= \frac{3c^2\phi^2 }{4\omega L^2},
\end{equation}
which is the well-known HDE density in the BD cosmology
\cite{Sheykhi1}. In the limit of Einstein gravity where $G_{\rm
eff}\rightarrow G$, then the BD scalar field becomes trivial, i.e.
$\phi^2=\omega/2\pi G=4\omega M_p^2$, and Eq. (\ref{rho1}) reduces
to the ECHDE density (\ref{rhoS}) in Einstein gravity \cite{wei}.

Following \cite{Huang}, the IR cut-off $L$ is defined as
\begin{equation}
L=a(t)\frac{\sin n\big(\sqrt{|k|}y\big)}{\sqrt{|k|}},\label{L}
\end{equation}
where
\begin{equation}
\frac{\sin n\big(\sqrt{|k|}y\big)}{\sqrt{|k|}}=\left\{
\begin{array}{ll}
\sin y,& k=1, \\
y,& k=0, \\
\sinh y,&k=-1, \\
\end{array}
\right.\label{sinn}
\end{equation}
and
\begin{equation}
y=\frac{R_{\rm h}}{a(t)}=\int_t^\infty \frac{{\rm d}
t}{a(t)}=\int_0^{r} \frac{{\rm d}r}{\sqrt{1-kr^2}}=\left\{
\begin{array}{ll}
\sin^{\rm -1}r,& k=1, \\
r,& k=0, \\
\sinh^{\rm -1}r,&k=-1. \\
\end{array}
\right.\label{y}
\end{equation}
Here $R_{\rm h}$ is the radial size of the event horizon measured in
the $r$ direction and $L$ is the radius of the event horizon
measured on the sphere of the horizon \cite{Huang}. For a flat
Universe, $L=R_{\rm h}$.

The critical energy density, $\rho_{\mathrm{cr}}$, and the energy
density of the curvature, $\rho_k$, are defined as
\begin{eqnarray}\label{rhocr}
\rho_{\mathrm{cr}}=\frac{3\phi^2 H^2}{4\omega},\hspace{0.8cm}
\rho_k=\frac{3k\phi^2}{4\omega a^2}.
\end{eqnarray}
The fractional energy densities are also defined as usual
\begin{eqnarray}
\Omega_M&=&\frac{\rho_M}{\rho_{\mathrm{cr}}}=\frac{4\omega\rho_M}{3\phi^2
H^2}, \label{Omegam} \\
\Omega_k&=&\frac{\rho_k}{\rho_{\mathrm{cr}}}=\frac{k}{H^2 a^2},\label{Omegak} \\
\Omega_D&=&\frac{\rho_D}{\rho_{\mathrm{cr}}}=\frac {c^2\gamma_c}{L^2H^2}.
\label{OmegaD}
\end{eqnarray}
For latter convenience we rewrite Eq. (\ref{OmegaD}) in the form
\begin{eqnarray}
HL=\left(\frac{c^2\gamma_c}{\Omega_{D}}\right)^{1/2}. \label{HL HDE
in BD}
\end{eqnarray}
Taking time derivative of Eq. (\ref{L}) and using (\ref{HL HDE in
BD}) yields
\begin{eqnarray}
\dot{L}=\left(\frac{c^2\gamma_c}{\Omega_{D}}\right)^{1/2}-\cos
n\big(\sqrt{|k|}y\big),\label{Ldot1}
\end{eqnarray}
where
\begin{equation}
\cos n\big(\sqrt{|k|}y\big)=\left\{
\begin{array}{ll}
\cos y,& k=1, \\
1,& k=0, \\
\cosh y,&k=-1. \\
\end{array}
\right.\label{cosn1}
\end{equation}
Using Eqs. (\ref{L}), (\ref{sinn}), (\ref{Omegak}) and (\ref{HL HDE
in BD}), one can rewrite Eq. (\ref{cosn1}) as
\begin{equation}
\cos
n\big(\sqrt{|k|}y\big)=\left[1-\Omega_k\left(\frac{c^2\gamma_c}{\Omega_D}\right)\right]^{1/2}.\label{cosn}
\end{equation}
Hence, Eq. (\ref{Ldot1}) yields
\begin{equation}
\dot{L}=\left(\frac{c^2\gamma_c}{\Omega_{D}}\right)^{1/2}\left[1-\left(\frac{\Omega_D}{c^2\gamma_c}
-\Omega_k\right)^{1/2}\right].\label{Ldot}
\end{equation}
For the FRW Universe containing the ECHDE and pressureless matter,
the continuity equations are
\begin{eqnarray}
&&\dot{\rho}_D+3H\rho_D(1+w_D)=0,\label{consq}\\
&&\dot{\rho}_M+3H\rho_M=0, \label{consm}
\end{eqnarray}
where $w_D=p_D/\rho_D$ is the EoS parameter of the ECHDE.

Taking the time derivative of Eq. (\ref{rho1}) and using
(\ref{dotphi}) and (\ref{Ldot}), we obtain
\begin{eqnarray}
\dot{\rho_D}=\Big(\frac{2H\rho_{D}}{\gamma_c}\Big)
\left\{2n\gamma_c+\left[1-2\gamma_c+\frac{4\omega}{\phi^2}\frac{\alpha
H^2}{3c^2}\Big(\frac{\Omega_{D}}{c^2\gamma_c}\Big)\right]
\left[1+n-\Big(\frac{\Omega_D}{c^2\gamma_c}-\Omega_k\Big)^{1/2}\right]\right\}.\label{ro
dot}
\end{eqnarray}
For $\alpha=0=\beta$ $(\gamma_c=1)$ we recover
\begin{eqnarray}
\dot{\rho}_D=2H\rho_D\left[n-1+\Big(\frac{\Omega_D}{c^2}-\Omega_k\Big)^{1/2}\right]\label{rhodot
HDE in B-D},
\end{eqnarray}
which is the same as the result obtained for the HDE in BD gravity
\cite{Sheykhi1}. While for $n=0$ and $\phi^2=4\omega M_P^2$, we have
\begin{eqnarray}
\dot{\rho_D}=\Big(\frac{2H\rho_{D}}{\gamma_c}\Big)
\left[1-2\gamma_c+\frac{\alpha
H^2}{3c^2M_P^2}\Big(\frac{\Omega_{D}}{c^2\gamma_c}\Big)\right]
\left[1-\Big(\frac{\Omega_D}{c^2\gamma_c}-\Omega_k\Big)^{1/2}\right].\label{ro
dot1}
\end{eqnarray}
Using Eqs. (\ref{rhoS}), (\ref{OmegaD}) and (\ref{HL HDE in BD}),
one can rewrite Eq. (\ref{ro dot1}) as
\begin{eqnarray}\label{rhodot ECHDE}
\dot{\rho}_D=\Big(\frac{c^2\gamma_c}{\Omega_{D}}\Big)^{1/2}\left[\frac{2\alpha-4\beta}{L^5}-\frac{4\alpha}{L^5}\ln(M_p^2
L^2)-\frac {6c^2
M_p^2}{L^3}\right]\left[1-\Big(\frac{\Omega_D}{c^2\gamma_c}-\Omega_k\Big)^{1/2}\right],
\end{eqnarray}
which is same as the result derived for the ECHDE in Einstein
gravity \cite{Jamil}. Substituting Eq. (\ref{ro dot}) in
(\ref{consq}) yields the EoS parameter of the ECHDE in BD gravity as
\begin{eqnarray}
w_D=-1-\frac{4n}{3}-\frac{2}{3\gamma_c}
\left[1-2\gamma_c+\frac{4\omega}{\phi^2}\frac{\alpha
H^2}{3c^2}\Big(\frac{\Omega_{D}}{c^2\gamma_c}\Big)\right]
\left[1+n-\Big(\frac{\Omega_D}{c^2\gamma_c}-\Omega_k\Big)^{1/2}\right].\label{wD1}
\end{eqnarray}
Note that as we already mentioned, at the very early stage when the
Universe undergoes an inflation phase, the correction terms in the
ECHDE density (\ref{rho1}) become important. After the end of the
inflationary phase, the Universe subsequently enters in the
radiation and then matter dominated eras. In these two epochs, since
the Universe is much larger, the entropy-corrected terms to ECHDE,
namely the last two terms in Eq. (\ref{rho1}), can be safely
ignored. During the early inflation era the Hubble parameter $H$ is
constant and $a=\exp{(Ht)}$. Hence, the Hubble horizon $H^{-1}$ and
the future event horizon $R_{\rm h}=a\int_t^\infty \frac{{\rm d}
t}{a}$ will coincide i.e. $R_{\rm h} = H^{-1}=$ const. On the other
hand, since an early inflation era leads to a flat Universe, i.e.
$\Omega_k=0$, we have $L = R_{\rm h} = H^{-1}=$ const. Also from Eq.
(\ref{HL HDE in BD}) we have $\frac{\Omega_{D}}{c^2\gamma_c}=1$.
Therefore during the early inflation era, Eq. (\ref{wD1}) reduces to
\begin{equation}
w_D=-1-\frac{2nc^2}{3\Omega_D}\left[1+\frac{4}{3}\frac{\alpha\omega}{c^2}\frac{H^2}{\phi^2}\right].\label{wDinf1}
\end{equation}
Using $\phi=a^n$, Eq. (\ref{wDinf1}) yields
\begin{equation}
w_D=-1-\frac{2nc^2}{3\Omega_D}\left[1+\frac{4}{3}\frac{\alpha\omega}{c^2}H^2e^{-2nHt}\right].
\label{wDinf2}
\end{equation}
It is worth while to mention that in BD gravity, besides the
standard de Sitter inflation ($a=\exp{(Ht)}$), two other
inflationary solutions namely the intermediate ($a=\exp{(At^f)}$,
$A$ and $f$ are constants) and the power-law ($a=t^p, p
> 1$) inflation can also be realized. In the intermediate
case, the expansion of the Universe is slower than de Sitter but
faster than power-law inflation \cite{Barrow}.

In the absence of correction terms $(\alpha=\beta=0)$ we have
$(\gamma_c=1)$ and Eq. (\ref{wD1}) recovers the EoS parameter of the
HDE in BD theory \cite{Sheykhi1}
\begin{eqnarray}
w_D=-\frac{1}{3}-\frac{2n}{3}-\frac{2}{3}\Big(\frac{\Omega_D}{c^2}-\Omega_k\Big)^{1/2}\label{wD2}.
\end{eqnarray}
Comparing Eq. (\ref{wDinf1}) with (\ref{wD2}) we see that in the
presence of correction terms the scalar field $\phi$ enters the EoS
parameter explicitly. From Eq. (\ref{wD1}) we see that when the
ECHDE is combined with BD field the transition from normal state
where $w_D >-1 $ to the phantom regime where $w_D <-1 $ for the EoS
of DE can be easily achieved. This is in contrast to Einstein
gravity where the EoS of noninteracting HDE cannot cross the phantom
divide $w_D=-1$ \cite{li}. To illustrate this result in ample
detail, we investigate it for the late-time Universe where
$\Omega_D=1$ and $\Omega_k=0$. In this case we have $L=R_{\rm h}\neq
H^{-1}$ and $H\neq$ const. Now from Eq. (\ref{wD1}) we find
$w_D=-\frac{1}{3}-\frac{2}{3c}-\frac{2n}{3}$. If we take $c=1$
\cite{Sheykhi1} then $w_D=-1-\frac{2n}{3}$. On the other hand for
ECHDE in Einstein gravity ($n\rightarrow0$) with $c=1$ we obtain
$w_D=-1$. Thus in the late-time Universe, although the EoS parameter
of ECHDE does not feel the presence of the last two correction terms
in Eq. (\ref{rho1}) but for $n\neq 0$ it will necessary cross the
phantom divide, i.e. $w_D<-1$ in BD theory. This is in contrast to
Einstein gravity ($n\rightarrow0$) where $w_D$ of ECHDE mimics a
cosmological constant in the late-time.

The deceleration parameter is given by
\begin{eqnarray}
q=-\frac{\ddot{a}}{aH^2}=-1-\frac{\dot{H}}{H^2}.
\end{eqnarray}
Dividing  Eq. (\ref{FE2}) by $H^2$, and using Eqs. (\ref{rho1}),
(\ref{HL HDE in BD}), (\ref{dotphi}) and (\ref{ddotphi}), we obtain
\begin{eqnarray}
q=\frac{1}{2n+2}\left[(2n+1)^2+2n(n\omega-1)+\Omega_k+3\Omega_D
w_D\right]\label{q1}.
\end{eqnarray}
Replacing $w_D$ from Eq. (\ref{wD1}), we obtain the deceleration
parameter for the ECHDE in BD theory as
\begin{eqnarray}
q&=&\frac{1}{2n+2}\left\{(2n+1)^2+2n(n\omega-1)+\Omega_k-(2n+1)\Omega_D-
2\Omega_D\Big(\frac{\Omega_D}{c^2\gamma_c}-\Omega_k\Big)^{1/2}\right.
\nonumber\
\\
&&
\left.-\frac{2\Omega_D}{\gamma_c}\left[\frac{4\omega}{\phi^2}\frac{\alpha
H^2}{3c^2}\Big(\frac{\Omega_{D}}{c^2\gamma_c}\Big)+1-\gamma_c\right]
\left[1+n-\Big(\frac{\Omega_D}{c^2\gamma_c}-\Omega_k\Big)^{1/2}\right]
\right\}\label{q2-nint}.
\end{eqnarray}
For $\alpha=\beta=0$ $(\gamma_c=1)$ Eq. (\ref{q2-nint}) reduces to
the case of HDE in BD gravity \cite{Sheykhi1}
\begin{eqnarray}
q&=&\frac{1}{2n+2}\left[(2n+1)^2+2n(n\omega-1)+\Omega_k-(2n+1)\Omega_D-
2\Omega_D\Big(\frac{\Omega_D}{c^2}-\Omega_k\Big)^{1/2}\right]\label{q3}.
\end{eqnarray}
If we take $\Omega_D= 0.73$ and $\Omega_k\approx 0.01$  for the
present time and choosing $c=1$, $n\omega\approx1$ and
$\omega=10^4$, we obtain $q\approx-0.48$ for the present value of
the deceleration parameter which is in good agreement with recent
observational results \cite{Daly}.
%%%%%%%%%%%%%%%%%%%%%%%%%%%%%%%%%%%%%%%%%%%%%%%%%%%%%%%%%%%%%%%%%%%%%%%%%%%%%%%%%%%%%%%%%%%%%%%%%%%%%%%%%%%%%%
\section{Interacting ECHDE in BD theory\label{INTECHDE}}
Here our aim is to extend our study for the case that there is an
interaction between ECHDE and DM. The recent observational evidence
provided by the galaxy cluster Abell A586 supports the interaction
between DE and DM \cite{Bertolami8}. This kind of interaction can be
detected in the formation of large scale structures. It was
suggested that the dynamical equilibrium of collapsed structures
such as galaxy clusters would get modification due to the coupling
between DE and DM \cite{Bertolami8}. The idea is that the virial
theorem is modified by the energy exchange between the dark sectors
leading to a bias in the estimation of the virial masses of clusters
when the usual virial conditions are employed. This provides a near
Universe probe of the dark coupling. The other observational
signatures on the dark sectors' mutual interaction can be found in
the probes of the cosmic expansion history by using the SNeIa,
baryonic acoustic oscillation (BAO), and CMB shift data, etc.
\cite{Guo}.

The total energy density satisfies the following conservation law
\begin{equation}\label{cons}
\dot{\rho}+3H(\rho+p)=0,
\end{equation}
where $\rho=\rho_M+\rho_D$ and $p=p_D$. Interaction causes the
ECHDE and DM do not conserve separately and they must rather enter
the energy balances \cite{Zimdahl}
\begin{eqnarray}
&&\dot{\rho}_D+3H\rho_D(1+w_D)=-Q,\label{consq2}\\
&&\dot{\rho}_{DM}+3H\rho_{DM}=Q, \label{consm2}\\
&&\dot{\rho}_{BM}+3H\rho_{BM}=0,\label{consbm}
\end{eqnarray}
where we have assumed the BM dose not interact with DE. Here $Q$
represents the interaction term. It is important to note that the
continuity equations imply that the interaction term should be a
function of a quantity with units of inverse of time (a first and
natural choice can be the Hubble factor $H$) multiplied with the
energy density. Therefore, the interaction term could be in any of
the following forms: (i) $Q\propto H \rho_D$, (ii) $Q\propto H
\rho_M$, or (iii) $Q\propto H (\rho_M+\rho_D)$. The more general
choice is $Q =3b^2 H(\rho_{M}+\rho_{D})$ with $b^2$ is a coupling
constant \cite{Ame}. The freedom of choosing the specific form of
the interaction term $Q$ stems from our incognizance of the origin
and nature of DE as well as DM. Moreover, a microphysical model
describing the interaction between the dark components of the
Universe is not available nowadays. Thus, in the absence of such a
theory, we rely on pure dimensional basis for choosing an
interaction $Q$.

Combining Eqs. (\ref{rhocr}) and (\ref{dotphi}) with the first
Friedmann equation (\ref{FE1}), we can rewrite this equation as
\begin{eqnarray}\label{rhos}
\rho_{\mathrm{cr}}+\rho_k=\rho_{BM}+\rho_{DM}+\rho_D+\rho_{\phi},
\end{eqnarray}
with the definition
\begin{eqnarray}\label{rhophi}
\rho_{\phi}\equiv\frac{1}{2}n H^2\phi^2\Big(n-\frac{3}{\omega}\Big).
\end{eqnarray}
Dividing Eq. (\ref{rhos}) by $\rho_{\rm cr}$, it can be rewritten
as
\begin{eqnarray}\label{Fried2new}
\Omega_{BM}+\Omega_{DM}+\Omega_D+\Omega_{\phi}=1+\Omega_k,
\end{eqnarray}
where
\begin{eqnarray}\label{Omegaphi}
\Omega_{\phi}=\frac{\rho_{\phi}}{\rho_{\mathrm{cr}}}=-2n
\left(1-\frac{n \omega}{3}\right).
\end{eqnarray}
By the help of the above definitions, one can rewrite the
interaction term as
\begin{eqnarray}\label{Q}
Q=3b^2H(\rho_{DM}+\rho_D)=3b^2H\rho_D(1+r),
\end{eqnarray}
where
\begin{eqnarray}\label{r}
r=\frac{\Omega_{DM}}{\Omega_D}=-1+\frac{1}{\Omega_D}\left[1+\Omega_k-\Omega_{BM}
+2n\left(1-\frac{n\omega}{3}\right)\right],
\end{eqnarray}
shows the ratio of the energy densities of two dark components.
Using the continuity equation (\ref{consbm}), one can easily obtain
\begin{equation}
\rho_{BM}=\rho_{BM_0}a^{-3}=\rho_{BM_0}(1+z)^3.\nonumber
\end{equation}
Dividing the above relation by $\rho_{\rm cr}=3\phi^2H^2/(4\omega)$
gives
\begin{equation}
\Omega_{BM}=\left(\frac{\rho_{\rm cr_0}}{\rho_{\rm
cr}}\right)\Omega_{BM_0}a^{-3}=\left(\frac{\rho_{\rm
cr_0}}{\rho_{\rm cr}}\right)\Omega_{BM_0}(1+z)^3,\nonumber
\end{equation}
where $\rho_{\rm cr_0}=3\phi_0^2H_0^2/(4\omega)$ and
 $\Omega_{BM_0}\sim 0.04$ is the fractional energy density of
baryoinc matter at the present time. Here $z=a^{-1}-1$ is the
cosmological redshift. Inserting Eqs. (\ref{ro dot}), (\ref{Q}) and
(\ref{r}) in (\ref{consq2}) we obtain the EoS parameter of the
interacting ECHDE in BD theory as
\begin{eqnarray}
w_D=-1-\frac{4n}{3}-\frac{2}{3\gamma_c}
\left[1-2\gamma_c+\frac{4\omega}{\phi^2}\frac{\alpha
H^2}{3c^2}\Big(\frac{\Omega_{D}}{c^2\gamma_c}\Big)\right]
\left[1+n-\Big(\frac{\Omega_D}{c^2\gamma_c}-\Omega_k\Big)^{1/2}\right]
\nonumber\\
-\frac{b^2}{\Omega_D}\left[1+\Omega_k-\Omega_{BM}+2n\left(1-\frac{n\omega}{3}\right)\right].\label{wDInt}
\end{eqnarray}
Comparing Eq. (\ref{wDInt}) with (\ref{wD1}) shows that in the
presence of interaction since the last expression in Eq.
(\ref{wDInt}) has a negative contribution, hence crossing the
phantom divide, i.e. $w_D<-1$, can be more easily achieved for than
when the interaction between ECHDE and DM is not considered.

During the early inflation era ($L = R_{\rm h} = H^{-1}=$ const.)
when the correction terms make sense in the ECHDE density
(\ref{rho1}), Eq. (\ref{wDInt}) yields
\begin{equation}
w_D=-1-\frac{2nc^2}{3\Omega_D}\left[1+\frac{4}{3}\frac{\alpha\omega}{c^2}H^2e^{-2nHt}\right]
-\frac{b^2}{\Omega_D}\left[1-\Omega_{BM}+2n\left(1-\frac{n
\omega}{3}\right)\right].\label{wDinfint}
\end{equation}
In the absence of correction terms $(\alpha=\beta=0)$, Eq.
(\ref{wDInt}) reduces to
\begin{eqnarray}
w_D=-\frac{1}{3}-\frac{2n}{3}-\frac{2}{3}\Big(\frac{\Omega_D}{c^2}-\Omega_k\Big)^{1/2}
-\frac{b^2}{\Omega_D}\left[1+\Omega_k-\Omega_{BM}+2n\left(1-\frac{n
\omega}{3}\right)\right]\label{wDInt1},
\end{eqnarray}
which is exactly the result obtained for the HDE in BD gravity
\cite{Sheykhi1}. On the other hand, when $n=0$
($\omega\rightarrow\infty$) the BD scalar field becomes trivial,
i.e. $\phi^2=\omega/2\pi G=4\omega M_P^2$, and Eq. (\ref{wDInt})
yields
\begin{eqnarray}
w_D=-1&-&\frac{2}{3\gamma_c}\left[1-2\gamma_c+\frac{\alpha
H^2}{3c^2M_P^2}\Big(\frac{\Omega_{D}}{c^2\gamma_c}\Big)\right]
\left[1-\Big(\frac{\Omega_D}{c^2\gamma_c}-\Omega_k\Big)^{1/2}\right]
~\nonumber\\&-&\frac{b^2}{\Omega_D}[1+\Omega_k-\Omega_{BM}].\label{wECHDE}
\end{eqnarray}
Using Eqs. (\ref{rhoS}), (\ref{OmegaD}) and (\ref{HL HDE in BD}),
one can rewrite Eq. (\ref{wECHDE}) as
\begin{eqnarray}\label{omegaD ECHDE }
w_D=-1&-&\left[\frac{(2\alpha-4\beta)L^{-2}-4\alpha L^{-2}\ln(M_p^2
L^2)-6c^2 M_p^2}{3\Big(3c^2M^2_p+\alpha L^{-2}\ln({M_p}^2 L^2)+\beta
L^{-2}\Big)}\right]\left[1-\Big(\frac{\Omega_D}{c^2\gamma_c}-\Omega_k\Big)^{1/2}\right]\nonumber\\&-&\frac{b^2}{\Omega_D}[1+\Omega_k-\Omega_{BM}],
\end{eqnarray}
which recovers its respective expression in ECHDE model in Einstein
gravity \cite{Jamil}. If we compare Eq. (\ref{wDInt}) with
(\ref{wECHDE}) we find out that when ECHDE is combined with BD field
the transition from normal state where $w_D >-1 $ to the phantom
regime where $w_D <-1 $ for the EoS of interacting DE can be more
easily achieved for than when resort to the Einstein field equations
is made.

Following \cite{Kim06}, if we define the effective EoS parameter
\begin{eqnarray}\label{wef}
w^{\mathrm{eff}}_D=w_D+\frac{\Gamma}{3H},
\end{eqnarray}
where $\Gamma=3b^2(1+r)H$. Then, the continuity equation
(\ref{consq2}) can be rewritten in the standard form
\begin{eqnarray}
&&\dot{\rho}_D+3H\rho_D(1+w^{\mathrm{eff}}_D)=0.\label{consqeff}
\end{eqnarray}
Substituting Eq. (\ref{wDInt}) into (\ref{wef}) yields
\begin{eqnarray}
w^{\mathrm{eff}}_D=-1-\frac{4n}{3}-\frac{2}{3\gamma_c}
\left[1-2\gamma_c+\frac{4\omega}{\phi^2}\frac{\alpha
H^2}{3c^2}\Big(\frac{\Omega_{D}}{c^2\gamma_c}\Big)\right]
\left[1+n-\Big(\frac{\Omega_D}{c^2\gamma_c}-\Omega_k\Big)^{1/2}\right].
\end{eqnarray}
For $\alpha=\beta=0$ then $\gamma_c=1$ and we have
\begin{eqnarray}\label{wDeff}
w^{\mathrm{eff}}_D=-\frac{1}{3}-\frac{2n}{3}-\frac{2}{3}\Big(\frac{\Omega_D}{c^2}-\Omega_k\Big)^{1/2},
\end{eqnarray}
which is same as the result obtained for HDE in BD theory
\cite{Sheykhi1}. It is important to note that in the literature,
``the effective EoS" is also defined as the EoS for the total energy
density $\rho_{\rm tot}$ and pressure $P_{\mathrm{tot}}$ of the
Universe which in the flat FRW Universe, is given by  $w_{\rm eff} =
-1 -2\dot{H}/\left(3H^2\right) =
P_{\mathrm{tot}}/\rho_{\mathrm{tot}}$ \cite{nojiri}. However, for
the interacting HDE models the effective EoS parameter is defined as
in Eq. (57) with adding the interaction term to $w_D$ \cite{Kim06}.
So this definition differs from that of \cite{nojiri}. From Eqs.
(59) and (60), one can easily see that $w_{\rm eff}$ in BD theory
can cross the phantom line provided the model parameters are chosen
suitably.

Substituting Eq. (\ref{wDInt}) into (\ref{q1}) yields the
deceleration parameter for the interacting ECHDE in BD gravity as
\begin{eqnarray}
q=\frac{1}{2n+2}\left\{(2n+1)^2+2n(n\omega-1)+\Omega_k-(2n+1)\Omega_D-2\Omega_D\Big(\frac{\Omega_D}{c^2\gamma_c}-\Omega_k\Big)^{1/2}\right. ~~~~~\nonumber\\
\left.-\frac{2\Omega_D}{\gamma_c}\left[\frac{4\omega}{\phi^2}\frac{\alpha
H^2}{3c^2}\Big(\frac{\Omega_{D}}{c^2\gamma_c}\Big)+1-\gamma_c\right]
\left[1+n-\Big(\frac{\Omega_D}{c^2\gamma_c}-\Omega_k\Big)^{1/2}\right]+3\Omega_D\zeta
\right\}\label{q2},
\end{eqnarray}
where
\begin{equation}
\zeta=-\frac{b^2}{\Omega_D}\left[1+\Omega_k-\Omega_{BM}+2n\left(1-\frac{n\omega}{3}\right)\right].
\end{equation}
In the absence of correction terms, i.e. $\alpha=\beta=0$, Eq.
(\ref{q2}) reduces to the deceleration parameter for the
interacting HDE in BD gravity \cite{Sheykhi1}
\begin{eqnarray}
q&=&\frac{1}{2n+2}\left\{(2n+1)^2+2n(n\omega-1)+\Omega_k-(2n+1)\Omega_D-2\Omega_D\Big(\frac{\Omega_D}{c^2}-\Omega_k\Big)^{1/2}\right.
\nonumber\
\\
&&
\left.-3b^2\left[1+\Omega_k-\Omega_{BM}+2n\left(1-\frac{n\omega}{3}\right)\right]\right\}\label{q2Int}.
\end{eqnarray}
We can also obtain the equation of motion for $\Omega_{D}$. Taking
time derivative of Eq. (\ref{OmegaD}) and using relation
${\dot{\Omega}_{D}}=H{\Omega'_{D}}$, we obtain
\begin{eqnarray}\label{OmegaD2}
{\Omega'_D}=2\Omega_D\left[q+\Big(\frac{\Omega_D}{c^2\gamma_c}-\Omega_k\Big)^{1/2}\right]~~~~~~~~~~~~~~~~~~~~~~~~~~~~~~~~~~~~~~~~~~~~\nonumber\\+\frac{2\Omega_D}{\gamma_c}\left[\frac{4\omega}{\phi^2}\frac{\alpha
H^2}{3c^2}\Big(\frac{\Omega_{D}}{c^2\gamma_c}\Big)+1-\gamma_c\right]\left[1+n-\Big(\frac{\Omega_D}{c^2\gamma_c}-\Omega_k\Big)^{1/2}\right],
\end{eqnarray}
where the prime denotes the derivative with respect to $x=\ln{a}$.
Also $q$ is given by Eq. (\ref{q2}). For $\alpha=\beta=0$, the above
expression reduces to the case of interacting HDE in BD gravity
\cite{Sheykhi1}
\begin{eqnarray}\label{OmegaD4}
{\Omega'_D}=\Omega_D\left\{(1-\Omega_D)\left[1+2\Big(\frac{\Omega_D}{c^2}-\Omega_k\Big)^{1/2}\right]
-3b^2(1+\Omega_k-\Omega_{BM})+\Omega_k\right\}.
\end{eqnarray}
%%%%%%%%%%%%%%%%%%%%%%%%%%%%%%%%%%%%%%%%%%%%%%%%%%%%%%%%%%%%%%%%%%%%%%%%%%%%%%%%%%%%%%%%%%%%%%%%%%%%%%%%%%%%%%
\section{Conclusions\label{CONC}}
In this paper, we investigated the model of HDE with the logarithmic
corrections. These corrections are motivated from the LQG which is
one of the promising theories of quantum gravity. We started by
taking a non-flat FRW background in the BD gravitational theory.
This theory involves a scalar field which accounts for a dynamical
gravitational constant. We assumed an ansatz by which the BD scalar
field evolves with the expansion of the Universe. We then
established a correspondence between the field and the ECHDE to
study its dynamics. The dynamics are governed by few dynamical
parameters like its EoS, deceleration and energy density parameters.
For the sake of generality, we calculated them in the non-flat
background with the interaction of ECHDE with the matter. The study
favors the phantom crossing scenario due to the availability of
abundant parameters. It is not our purpose to fix or fit these
parameters and we left it till the availability of the observational
data. We hope that the future high precision observations like the
type Ia supernovae (SNeIa) surveys, the shift parameter of the
cosmic microwave background (CMB) observed by the Wilkinson
Microwave Anisotropy Probe (WMAP) and the Planck Mission, and the
baryon acoustic oscillation (BAO) measurement from the Sloan Digital
Sky Survey (SDSS) may be capable for determining the fine property
of the interacting entropy-corrected holographic model of DE in BD
gravity and consequently reveal some significant features of the
underlying theory of DE.

%%%%%%%%%%%%%%%%%%%%%%%%%%%%%%%%%%%%%%%%%%%%%%%%%%%%%%%%%%%%%%%%%%%%%%%%%%%%%%%%%%%%%%%%%%%%%%%%%%%%%%%%%%%%%%
\acknowledgments{The works of A. Sheykhi and K. Karami have been
supported financially by Research Institute for Astronomy and
Astrophysics of Maragha (RIAAM), Maragha, Iran.  M. Jamil would like
to thank the Abdus Salam International Center for Theoretical
Physics (ICTP), Trieste, Italy for its kind hospitality during which
part of this work was completed. Useful comments from the anonymous
referee are also gratefully acknowledged.}
%%%%%%%%%%%%%%%%%%%%%%%%%%%%%%%%%%%%%%%%%%%%%%%%%%%%%%%%%%%%%%%%%%%%%%%%%%%%%%%%%%%%%%%%%%%%%%%%%%%%%%%%%%%%%%

%%%%%%%%%%%%%%%%%%%%%%%%%%%%%%%%%%%%%%%%%%%%%%%%%%%%%%%%%%%%%%%%%%%%%%%%%%%%%%%%%%%%%%%%%%%%%%%%%%%%%%%%%%%%%%
\end{document}